\documentstyle[preprint,aps,pra]{revtex}

\begin{document}

\draft

\title{Motional narrowing effect in one-dimensional Frenkel chains with 
configurational disorder}

\author{V.\ A.\ Malyshev$^{\dag}$ and F.\ Dom\'{\i}nguez-Adame}
\address{GISC, Departamento de F\'{\i}sica de Materiales, Universidad
Complutense, E-28040 Madrid, Spain}

\date{\today}

\maketitle

\begin{abstract}

We analyze the peculiarities of the motional narrowing effect in disordered
one-dimensional Frenkel chains with off-diagonal disorder, induced by
uncorrelated Gaussian fluctuations in the positions of the host  units. A clear
difference in the scaling laws with respect to the  magnitude of the positional
disorder and the chain size is found, as compared to those for uncorrelated
diagonal disorder as well as for off-diagonal disorder modeled by uncorrelated
randomness in the  nearest-neighbor couplings. The origin of such a difference
is discussed in detail.

\end{abstract}

\pacs{PACS number(s):
   71.35.Aa;   % Frenkel excitons and self-trapped excitons
   36.20.Kd;   % Electronic structure and spectra
   78.30.Ly    % Disordered solids
}

\section{Introduction}

The concept of motional narrowing, earlier raised by Knapp~\cite{Knapp84} for
the on-site energy (or diagonal) disorder, has been very fruitful in explaining
many optical phenomena in quasi-one-dimensional systems like J~aggregates of
polymethine dyes and conjugated polymers (for a review, see
Refs.~\cite{Fidder93,Spano94,Knoester96} and references therein). This effect
is manifested as a decrease of the magnitude of the diagonal disorder  as soon
as the states of individual molecules are collectivized due to the
intermolecular interaction and form excitonic states.  The decrease depends on
whether or not the disorder is small enough to be regarded as a perturbation.
In the perturbative case, the suppression factor is determined by the square
root of the whole number of molecules in an aggregate, while in the
nonperturbative case this number should be substituted by the number of
coherently bound molecules~\cite{Knapp84}. A similar narrowing effect, but
different suppression factor, was recently reported for the case of dynamic 
disorder~\cite{Malyshev98,Wubs98,Bakalis99}.

It was found in Ref.~\cite{Tilgner90} that the numerically simulated 
absorption spectrum of polysilane with an uncorrelated Gaussian  distribution 
of nearest-neighbor couplings are similar to those for  an uncorrelated
diagonal disorder (see also~\cite{Kozlov98}). In contrast, numerical
simulations of off-diagonal disorder given by Gaussian randomness in the
molecular positions~\cite{Fidder93,Fidder91} found that the different behavior 
of the optical observables was very different from that expected from standard
motional narrowing arguments~\cite{Knapp84}. 

The main goal of the present Letter is to uncover the origin of such a
difference. It will be shown that, in the case of configurational disorder,
certain correlations appear in the distribution of hopping integrals, in spite
of the fact that the distribution of the molecular positions is uncorrelated.
This finally results in a scaling law for the typical fluctuation of the
Frenkel Hamiltonian matrix with respect to the magnitude of positional disorder
and the number of molecules different from that for uncorrelated diagonal
disorder. As a consequence, the main features of the exciton optical response
are largely affected.

\section{Description of the model}
\label{model}

Consider a collection of $N$ two-level molecules forming a one-dimensional
lattice. For our present purposes, we will neglect the static inhomogeneous
offset energy of the molecules imposed by the surrounding host medium (diagonal
disorder). Under this assumption, the effective Frenkel Hamiltonian describing
the system can be written in the form  
\begin{equation}
{\cal H}= 
\sum_{nm} \>J_{nm}^{}|n\rangle \langle m| \ .
\label{Hamiltonian}
\end{equation}
Here, the state vector $|n\rangle$ denotes the $n$-th molecule being excited
and the site index $n$ lies within the symmetric domain $-(N-1)/2 \le n \le
(N-1)/2$ ($N$ assumed to be odd). $J_{nm}$ is the intersite interaction, which
will be considered to be of dipole origin, so that $J_{nm}= J/|n-m
+\xi_{n}-\xi_m|^3$, $n\neq m$, and $J_{nn}\equiv 0$; $\xi_n$ is the deviation
of the $n$-th molecule from its regular  position, which we  suppose that it
occurs only along the directions toward the two adjacent molecules. This
restriction allows us to replace the fluctuations of vector positions by
scalars. We note that, arguably, this assumption does not affect the
conclusions and is introduced only to simplify the analytical treatment. The
distribution function of $\xi_n$ is chosen to be Gaussian,
\begin{equation}
P(\xi_n)={1\over \sigma \sqrt{2\pi}}
        \exp\left( -{\xi_n^2 \over 2\sigma^2} \right)\ ,
\label{P}
\end{equation}
with variance $\sigma^2$.

To study the motional narrowing effect, one  should rewrite the Hamiltonian
(\ref{Hamiltonian}) in the excitonic representation using the eigenfunctions of
the unperturbed Hamiltonian (with no disorder). For the sake of simplicity, we
assume periodic  boundary conditions. Then, Bloch plane waves are the proper 
eigenfunctions of (\ref{Hamiltonian}) in the absence of disorder: 
\begin{equation}
|K\rangle = {1\over\sqrt N} \sum_{n}
e^{iKn}|n\rangle \ ,
\label{K}
\end{equation}
where $K=2\pi k/N$ belongs to the first Brillouin zone ($-(N-1)/2 \le k \le
(N-1)/2$). 

In the $K$-representation, the Hamiltonian (\ref{Hamiltonian}) reads
\begin{mathletters}
\label{1}
\begin{eqnarray}
{\cal H} & = & \sum_K E_K |K\rangle \langle K| + \sum_{K,K^\prime}
\Delta_{KK^\prime} |K\rangle \langle K^\prime| \ , \label{Hex}   \\
E_K & = & 2J \sum_{n>0} \> {1\over n^3}\cos (Kn) \ , \label{E_K} \\
\Delta_{KK^\prime} & = & {1\over N} \sum_{mn} \> \delta J_{nm}
e^{i(Kn - K^\prime m)} \ , \label{DeltaKK'} \\
\delta J_{nm} &=& J\left({1\over |n-m + \xi_n -\xi_m|^3} - 
{1\over |n-m|^3}\right) \ , \label{delta_J}
\end{eqnarray}
\end{mathletters}
\noindent where the scattering matrix $\Delta_{KK^\prime}$ is real due to the
symmetry of the summation region in (\ref{DeltaKK'}). Its diagonal part gives
rise to exciton energy shift, yielding inhomogeneous broadening of exciton
levels for an ensemble of chains. This effect appears to be the main
consequence of the presence of disorder if the off-diagonal term can be
regarded as a perturbation. The typical fluctuation of $\Delta_{KK}$ has a
direct relation with the inhomogeneous width of the corresponding exciton state
$|K\rangle$.

For nonperturbative magnitudes of the disorder, the off-diagonal elements
$\Delta_{KK^\prime}$ ($K\ne K^\prime$) mix the exciton states, resulting in
their localization on chain segments of a typical size smaller than the chain
length and subsequently affecting the exciton optical response.  Recall that,
for a perfect chain, only the state with $K=0$ is coupled to the light and
carries the entire exciton oscillator strength, which is then $N$ times larger
than that for an isolated molecule. Being mixed with other (nonradiative)
states ($K\neq 0$), the radiative state loses a part of the oscillator strength
due to its spreading over to the nonradiative ones. Thus, an effective number,
usually called number of coherently bound molecules $N^*<N$, replaces the
system size as the enhancement factor of the oscillator strength of the
localized exciton states~\cite{Knapp84}. It reflects the typical number of
sites on which the localized exciton wave functions have a significant
magnitude or, in other words, the number of molecules within a typical
localization segment. Accordingly, the inhomogeneous width of the optical
exciton line will also be subjected to renormalization~\cite{Knapp84}. In the
next Section, we will formulate a condition separating the  perturbative and
nonperturbative ranges of disorder magnitudes in  order to treat the exciton
optical response.

\section{Motional narrowing effect}
\label{MNE}

To gain insight into the magnitude of the typical fluctuation of the
scattering matrix $\Delta_{KK^\prime}$, one should calculate either its
distribution function or its moments, using the distribution~(\ref{P}) of
the positional fluctuations. We chose the second  procedure,  so that the
magnitude of interest will be the mean square deviation, defined as
\begin{equation}
\sigma_{KK^\prime}^2 = 
\Big\langle \Delta_{KK^\prime}^2\Big\rangle -
\Big\langle \Delta_{KK^\prime}\Big\rangle ^2\ ,
\label{sigma^2}
\end{equation}
where brackets denote the average over the joint probability distribution
$\prod_{n} P(\xi_n)$, with $P(\xi_n)$ of the form~(\ref{P}). 

In what follows, we assume that the standard deviation $\sigma$ is small.
Accounting then for that 
$$
{1\over |n-m + \xi_n -\xi_m|^3} = {1\over |n-m|^3}
\left[1 + {2(n - m)(\xi_n -\xi_m)\over (n-m)^2} + {(\xi_n -\xi_m)^2
\over (n - m)^2}\right]^{-3/2}\ ,
$$
one can expand $\delta J_{nm}$ in Eq.~(\ref{delta_J}) in Taylor series up
to fourth order with respect to 
$$
X_{nm} \equiv {2(n - m)(\xi_n -\xi_m) + (\xi_n -\xi_m)^2
\over (n - m)^2}\ ;
$$
we are interested in the contribution to $\sigma_{KK^\prime}^2$ up to the 
same order. The corresponding expressions for $\Delta_ {KK^\prime}$ and
$\Delta_ {KK^\prime}^2$ read
\begin{mathletters}
\label{1A}
\begin{eqnarray}
\Delta_ {KK^\prime} & = & {J\over N} \sum_{mn} {e^{i(Kn - K^\prime m)}
\over |n-m|^3}\left(-{3\over 2} X_{nm} + {15\over 8} X_{nm}^2 - 
{35\over 16} X_{nm}^3 + {315\over 128} X_{nm}^4 \right)\ , \\
& & \nonumber\\
\Delta_{KK^\prime}^2 & = & \left({J\over N}\right)^2 
\sum_{mn} {e^{i(Kn - K^\prime m)}\over |n-m|^3} 
\sum_{pq} {e^{-i(Kq - K^\prime p}) \over |q-p|^3}  \nonumber \\
& & \nonumber\\
&\times & \left({9\over 4} X_{nm}X_{qp}  
- {45\over 8} X_{nm}X_{qp}^2 
+ {105\over 16} X_{nm}X_{qp}^3 
+ {225\over 64} X_{nm}^2X_{qp}^2 \right)\ .
\end{eqnarray}
\end{mathletters}
Carrying out the average in Eqs. (\ref{1A}), we will collect all terms up to
fourth order in $\sigma$. The calculations are rather tedious  but
straightforward, so we only quote the final results:
\begin{mathletters}
\label{2A}
\begin{eqnarray}
\langle \Delta_ {KK^\prime}\rangle 
& = & J \delta_{KK^\prime}\Big[12JQ_5(K)\sigma^2 + 180Q_7(K) \sigma^4\Big]\ ,\\
& & \nonumber\\
\langle \Delta_{KK^\prime}^2\rangle 
& = & {9J^2\sigma^2\over N}\Big |P_5(K)+P_5^*(K^\prime)\Big |^2 
+ 144J^2\sigma^4 Q_5^2(K)\delta_{KK^\prime}\nonumber\\
& & \nonumber \\
& + & {360J^2\sigma^4\over N}\Big [P_5(K)+P_5^*(K^\prime)\Big]
      \Big[P_7^*(K)+P_7(K^\prime)\Big ] \nonumber \\
& & \nonumber \\
&+& {72J^2\sigma^4\over N}\left[ \Big[ Q_5(K)+
Q_5(K^\prime)\Big]^2 + 2\Big[ Q_{10}(0)+ Q_{10}(K+K^\prime)\Big]\right]\ ,
\end{eqnarray}
\end{mathletters}
\noindent where the functions $Q_\nu (K)$ and $P_\nu (K)$ are given by
\begin{equation}
P_\nu (K) = \sum_{m} {m e^{iKm} \over |m|^\nu}\ , \quad \quad
Q_\nu (K) = \sum_{m} {e^{iKm} \over |m|^\nu}\ , 
\label{PQ}
\end{equation}
and the term $m=0$ is excluded in the summations.

For the magnitude of interest, one obtains
\begin{eqnarray}
\sigma_{KK^\prime}^2 & = &
{9J^2\sigma^2\over N}\Big |P_5(K)+P_5^*(K^\prime)\Big |^2 
+ {360J^2\sigma^4\over N}\Big [P_5(K)+P_5^*(K^\prime)\Big]
\Big[P_7^*(K)+P_7(K^\prime)\Big ] \nonumber\\
& & \nonumber\\ 
& + & {72J^2\sigma^4\over N}\left[ \Big[ Q_5(K)+
Q_5(K^\prime)\Big]^2 + 2 \Big[ Q_{10}(0)+
Q_{10}(K+K^\prime)\Big]\right]\ .
\label{sigma_gen}
\end{eqnarray}
We stress that only terms up to fourth order in $\sigma$ are kept. 

To treat the optical properties of linear aggregates, the important wavenumbers
$K$ and $K^\prime$ are the smaller ones, namely $|K|,|K|^\prime \ll 1$. In this
long wavelength limit, the functions $P_\nu(K)$ and $Q_\nu (K)$ can be
approximated by their values within the nearest-neighbor framework, namely
$P_\nu (K) \approx 2iK$ and $Q_\nu (K) \approx 2$, accounting in the
sums~(\ref{PQ}) only for leading terms with $m = \pm 1$.
Equation~(\ref{sigma_gen}) then reduces to
\begin{equation}
\sigma_{KK^\prime}^2 ={36J^2\over N} (K-K^\prime)^2 \sigma^2 
(1+40 \sigma^2) + {1728J^2\over N}\sigma^4 \ .
\label{sigma_part}
\end{equation}
Now it can be clearly seen why one should keep the terms up to fourth order in
$\sigma$. The first term, being of second order with respect to $\sigma$, is
equal to zero for the diagonal elements of $\sigma_{KK^\prime}$ [note that it
is true independently of the magnitude of $K$ since $P_\nu(K)+P_\nu^*(K)\equiv
0$]. Thus, for $\sigma_{KK}^2$ one gets
\begin{equation}
\sigma_{KK}^2 = {1728J^2\over N}\sigma^4 \ .
\label{sigma_part_diag}
\end{equation}
This result differs from that for uncorrelated diagonal disorder, where the
corresponding magnitude scales as $\sigma^2/N$, $\sigma^2$ being the variance
of the site energy  distribution~\cite{Knapp84}. The same behavior appears as
well when one simulates off-diagonal disorder by uncorrelated randomness in the
nearest-neighbor hopping integrals~\cite{Tilgner90,Kozlov98} (here, $\sigma^2$
stands for the variance of the corresponding distribution). 

With respect to the off-diagonal elements of $\sigma_{KK^\prime}^2$, we  should
note that, in spite of the fact that the first term scales as $\sigma^2$, it
has an additional suppression factor proportional to $(K-K^\prime)^2 \sim
N^{-2}$, and thus it may be smaller than the fourth order one. When the first
term in  Eq.~(\ref{sigma_part}) dominates $\sigma_{KK^\prime}^2 \propto
\sigma^2/N^3$, while in the opposite case   one has $\sigma_{KK^\prime}^2
\propto \sigma^4/N$. Both results also differ from the scaling law $\sigma^2/N$
found for the other types of disorder~\cite{Knapp84,Tilgner90,Kozlov98}.

The origin of the difference discovered lies in that the terms linear in
$\xi_n$ in the fluctuations of the dipolar coupling of site $n$ to the
adjacent ones have the same magnitude but opposite signs, thus appearing to be
correlated, notwithstanding the fact that the fluctuations of  molecular
positions are completely uncorrelated. Due to this feature, they almost (or
exactly at $K=K^\prime$) cancel each other in~(\ref{DeltaKK'}) when summing
over $n$ in the long wavelength limit ($|K|, |K^\prime|\ll 1$). 

As it was noted in the previous Section, for perturbative magnitudes of
disorder, the main effect of $\sigma_{KK^{\prime}}$ is the broadening of the
exciton levels. The  value of $\sigma_{KK}$ given by
Eq.~(\ref{sigma_part_diag}) gives the half width of the $K$-th exciton state
and will serve for determining the latter provided that $\sigma_{KK^\prime}$
remains smaller than the corresponding energy differences $|E_K -
E_{K^\prime}|$. Since the minimum energy difference in the exciton spectrum is
between the state with $K = 0$ and the next one with $K=2\pi/N$, the equality 
\begin{equation}
\sigma_{K=2\pi/N,\ K^\prime=0} = |E_{K=2\pi/N} - E_{K^\prime=0}|
\label{threshold}
\end{equation}
determines a value of $\sigma$ which separates the ranges of perturbative
and nonperturbative magnitudes of disorder. In~(\ref{threshold}), $E_K$ is 
given by Eq.~(\ref{E_K}).

\section{Discussion of the numerical results}
\label{Discus}

As it was already mentioned in the Introduction, numerical simulations of
optical properties of linear molecular aggregates with off-diagonal disorder
generated by Gaussian uncorrelated fluctuations in the molecular positions
yielded different behaviors of the optical observables as compared to those for
diagonal disorder~\cite{Fidder93,Fidder91}.   In this Section, it will be shown
that the peculiarities found in Refs.~\cite{Fidder93,Fidder91} can be
qualitatively explained from the viewpoint of the modified motional narrowing
formula~(\ref{sigma_part}).   In particular, we will focus on the dependence of
the absorption band width $\sigma^*$ and the radiative rate enhancement factor 
on the degree of disorder $\sigma$. The radiative rate enhancement factor is
proportional to the number of coherently bound molecules $N^*$, while
$\sigma^*$ can be estimated from $\sigma_{KK}$ replacing $N$ by
$N^*$~\cite{Knapp84,Malyshev93}.  Thus, the magic number $N^*$ is, in fact, 
the unique quantity determining the observables we are interested in and,
correspondingly, our main goal is to obtain the dependence of $N^*$ on
$\sigma$. To do that, we will follow a simple rule earlier proposed in
Ref.~\cite{Malyshev91} (see also~\cite{Malyshev93}) that works surprisingly
well for explaining the corresponding data for diagonal
disorder~\cite{Malyshev93,Malyshev91}. This rule simply consists of applying
the formula~(\ref{threshold}) to the typical localization segment of size
$N^*$. In our estimates of $N^*$, we will keep only the second term in
Eq.~(\ref{sigma_part}) since, as will be shown later, it is the major
contribution for the parameters used in Refs.~\cite{Fidder93,Fidder91}.  

As a first step, let us take the exciton energy spectrum in the 
nearest-neighbor approximation: $E_K = 2J\cos K \approx 2J - JK^2$. After
these simplifications, we arrive at the following formula for the number of
coherently bound molecules
\begin{equation}
N^* = \left({\pi^4\over 108}\right)^{1/3} \sigma^{-4/3}
\approx  \sigma^{-4/3}\ . 
\label{N^*}
\end{equation}
Estimating now the absorption band width $\sigma^*$ as $2\sigma_{KK}$ by
replacing $N$ by $N^*$, we find
\begin{equation}
\sigma^* = 2\sqrt{1728}\> J \sigma^{8/3} \approx 83J \sigma^{8/3}\ . 
\label{sigma^*}
\end{equation}
Recall that for diagonal disorder the respective quantities scale as 
$\sigma^{-2/3}$~\cite{Malyshev93,Malyshev91} and 
$\sigma^{4/3}$~\cite{Fidder93,Fidder91,Malyshev93,Malyshev91,Schreiber81,%
Boukahil90,Kohler89}, whenever the nearest-neighbor approximation is  adopted.
It is worth noting that the new scaling laws~(\ref{N^*}) and~(\ref{sigma^*})
follow from the latters ($\sigma^{-2/3}$ and $\sigma^{4/3}$) simply replacing
$\sigma$ by $\sigma^2$, reflecting the fact that in our case the effective
disorder scales  as $\sigma^2$ instead of $\sigma$ [see
Eq.~(\ref{sigma_part_diag})]. 

The authors of Refs.~\cite{Fidder93,Fidder91} did not restrict themselves to
the nearest-neighbor approximation but took into account all dipolar couplings.
Using the parameterization $c\sigma^\alpha$, they found that  their numerical
data for  the factor of radiative rate enhancement and $\sigma^*$ were fitted
by the sets $c = 0.20$, $\alpha = -1.64$ and $c = 425J$, $\alpha = 2.84$,
respectively. The corresponding exponents for the case of diagonal disorder
were found to be $-0.74$ and $1.34$, respectively. Here, one also observes
approximately the two-times increase of the exponents when passing from
diagonal to off-diagonal disorder. This feature unambiguously shows that the
effective degree of disorder  in the last case scales as $\sigma^2$, what
perfectly correlates with our finding given by Eq.~(\ref{sigma_part_diag}).

Comparison of the numerical fits of Refs.~\cite{Fidder93,Fidder91} with  our
results~(\ref{N^*}) and~(\ref{sigma^*}), obtained under the assumption  of
nearest-neighbor coupling, shows that the numerical $\sigma$-scaling of  both
quantities is reproduced reasonably well by the theoretical estimates, better
for the absorption band width and worse for the number of  coherently bound
molecules.  Similar peculiarities are present in the  case of diagonal disorder
(compare $4/3$ with $1.34$ and $-2/3$ with $-0.74$). As it was firstly
mentioned in Refs.~\cite{Fidder93,Fidder91}, including all dipolar couplings
affects largely the factor of  radiative rate enhancement, rising it by more
than a factor $2$ as compared to that calculated in the nearest-neighbor
approximation. The explanation was done in Ref.~\cite{Malyshev95} and was based
on the exact exciton energy spectrum close to the bottom of the band
\begin{equation}
E_K = -2J\xi(3) + JK^2 \left({3\over 2} - \ln K \right) \ ,
\label{Ek_exact}
\end{equation}
where $\xi(3) = \sum_{n=1}^\infty n^{-3} = 1.202$. Notice that the 
presence of a logarithmic term in this equation results in a larger 
energy level separation as compared to the case of the nearest-neighbor 
model. Thus, a smaller number of the exciton states will be effectively 
mixed by disorder giving finally rise to an increase of $N^*$. Indeed, if 
we now use Eq.~(\ref{Ek_exact}) for finding $N^*$ we then arrive at the 
equation
\begin{equation}
{N^*\sqrt{N^*} \over \ln{N^*} - \ln{(2\pi) + 3/2 }}
= {\pi^2\over 6\sqrt{3}\sigma^2}  \ .
\label{N*_exact}
\end{equation}
For the nearest-neighbor model, one should substitute the denominator in
the left hand side of Eq.~(\ref{N*_exact}) by unity, getting then the 
previous expression for $N^*$, Eq.~(\ref{N^*}). At a fixed magnitude of 
$\sigma$, keeping the denominator will increase the value of $N^*$. 

It should be stressed that, rigorously speaking, from Eq.~(\ref{N*_exact}) 
does not follow a power-like behavior of $N^*$ against $\sigma$. Therefore,
we tried to fit the numerical data of Refs.~\cite{Fidder93,Fidder91} 
relative to the factor of radiative rate enhancement and the absorption 
band  width using the following parameterization
\begin{mathletters}
\label{3}
\begin{equation}
{N^*\sqrt{N^*} \over \ln{N^*} - a}
= {b \over \sigma^2}  \ ,
\label{N*_param}
\end{equation}
\begin{equation}
\sigma^* = c\> {J \sigma^2 \over \sqrt{N^*}}\ . 
\label{sigma*_param}
\end{equation}
\end{mathletters}
\noindent
The fits were reached provided $a = 0.004$, $b =0.12$ and $c =148$. Note that
the numerical factor in~(\ref{sigma*_param}) ($c =148$) does not differ
drastically from the theoretical value $48\sqrt{3} \approx 83$.

Finally, we would like to show that for disorder in the interval $0.025 <
\sigma < 0.08$, used in the numerical simulations~\cite{Fidder93,Fidder91}, the
first term in Eq.~(\ref{sigma_part}), neglected by us when making estimates, is
smaller than the second one.  Their ratio at $K^\prime=0$ and $K=2\pi/N$ is
equal to $\pi^2(1+40\sigma^2)/(12N^2\sigma^2)$. Substituting here $N$ by $N^*
\approx \sigma^{-4/3}$, we indeed get for this ratio a small magnitude
$\pi^2(1+40\sigma^2)\sigma^{2/3}/12 \ll 1$ when $\sigma$ ranges within the
above mentioned interval.  Thus, our assumption is self-consistent.

\section{Conclusion}
\label{Concl}

The motional narrowing effect in one-dimensional Frenkel chains with
off-diagonal disorder arising from Gaussian fluctuations in the molecular
positions is found to be different from that for diagonal disorder and for
off-diagonal disorder with uncorrelated randomness in the nearest-neighbor
couplings. Such distinction is due to the fact that the fluctuations of the
dipolar coupling of a given molecule to the adjacent ones are correlated, even
if the fluctuations of the molecular positions are completely uncorrelated.
Thus, this type of disorder cannot be modeled by uncorrelated randomness in the
nearest-neighbor interactions. The estimates of scaling of the optical
observables with the degree of disorder based on the new motional narrowing law
are found to be in qualitative agreement with those obtained previously in
numerical simulations.

\acknowledgments

The authors thank A.\ S\'{a}nchez for critical reading of the manuscript.
This work was supported by CAM under Project 07N/0034/98.  V.\ A.\ M.\
thanks UCM for the support under {\em Sab\'aticos Complutense\/} as well
as partial support from the Russian Foundation for Basic Research under
Project 97-03-09221.

\end{document}